\documentclass[prd,preprint,tightenlines,floatfix,showpacs,preprintnumbers,nofootinbib,eqsecnum]{revtex4}

 \usepackage[dvips,final]{graphicx}
  \usepackage{amssymb}
   \usepackage{amsmath}
    \usepackage{amsfonts}
     \usepackage{epsfig}
      \usepackage{bm}

\begin{document}

\thispagestyle{empty} \preprint{\hbox{}} \vspace*{-10mm}

\title{Elastic double diffractive production \\
of axial-vector $\chi_c(1^{++})$ mesons
and the Landau-Yang theorem}

\author{R.~S.~Pasechnik}
\email{rpasech@theor.jinr.ru}
\affiliation{Department of Physics, Arizona State University,
Tempe, AZ 85287-1504, USA and\\
Bogoliubov Laboratory of Theoretical Physics, JINR,
Dubna 141980, Russia}

\author{A.~Szczurek}
\email{antoni.szczurek@ifj.edu.pl}
\affiliation{Institute of Nuclear Physics PAN, PL-31-342 Cracow,
Poland and\\
University of Rzesz\'ow, PL-35-959 Rzesz\'ow,
Poland}

\author{O.~V.~Teryaev}
\email{teryaev@theor.jinr.ru}
\affiliation{Bogoliubov Laboratory of Theoretical Physics, JINR,
Dubna 141980, Russia}

\date{\today}

\begin{abstract}
We discuss exclusive elastic double diffractive axial-vector
$\chi_c(1^{+})$ meson production in proton-antiproton collisions at
the Tevatron. The amplitude for the process is derived within the
$k_t$-factorisation approach with unintegrated gluon distribution
functions (UGDFs). We show that the famous Landau-Yang theorem is
not applicable in the case of off-shell gluons. Differential cross
sections for different UGDFs are calculated. We compare exclusive
production of $\chi_c(1^+)$ and $\chi_c(0^+)$. The contribution of
$\chi_c(1^+)$ to the $J/\Psi + \gamma$ channel is smaller than that
of the $\chi_c(0^+)$ decay, but not negligible and can be measured.
The numerical value of the ratio of the both contributions is almost
independent of UGDFs modeling.
\end{abstract}


\maketitle

\section{Introduction}

The central exclusive production of mesons has been
recently revived. This is essentially because of
two reasons.

Firstly, the theoretical QCD inspired approach has been developed.
This is because of interest in the double diffractive production of
the Higgs boson firstly proposed by A. B. Kaidalov, V. A. Khoze, A.
D. Martin and M. G. Ryskin \cite{KMR,KKMR,KKMR-spin} (KKMR) as an
alternative to inclusive production for Higgs searches. In
principle, very similar methods can be used for scalar \cite{PST07},
pseudoscalar \cite{SPT07}, axial-vector and tensor mesons. The
situation for vector meson production is somewhat different. Here
the dominant mechanism is photon-pomeron (pomeron-photon) fusion
\cite{SS07} or pomeron-odderon (odderon-pomeron) fusion
\cite{BMSC07}. Recently a $k_t$-factorization approach has been used
to calculate exclusive $\Upsilon$ production \cite{RSS08} at
Tevatron.

Secondly, some experimental efforts have been done to facilitate
real measurements at Tevatron \cite{Pinfold_talks} and in the future
at RHIC \cite{Guryn} and LHC \cite{Schicker}. Some preliminary
results from Tevatron have been presented recently
\cite{Pinfold_talks,Albrow_diffraction2008}.
\begin{figure}[!h]    
 \centerline{\includegraphics[width=0.4\textwidth]{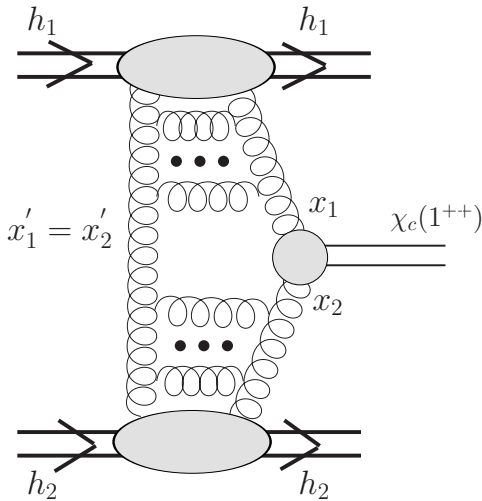}}
   \caption{\label{fig:QCDdiff}
   \small  The sketch of the bare QCD mechanism for diffractive
production of the $\chi_c(1^{+})$ meson. Some kinematical variables
are shown in addition.}
\end{figure}

In the present paper we are concentrated on the exclusive
double-diffractive production of axial-vector $\chi_c(1^{+})$
mesons\footnote{Some general aspects of this process were discussed
previously in Refs.~\cite{KKMR-spin,Close,Yuan01}.}. Here the
dominant mechanism is the two-pomeron fusion, which in the QCD
language is a ``fusion'' of two QCD ladders. The mechanism is shown
schematically in Fig.~\ref{fig:QCDdiff}. Compared to the Higgs
production, where a hard scale is guaranteed by the large mass of
the Higgs boson, here the natural scale (mass of the $\chi_c(1^+)$
meson) is much lower and the method proposed by KKMR is a bit
questionable as a big part of the strength may come from the region
of relatively small gluon transverse momenta. A pragmatic solution
is to use nonperturbative models of UGDFs instead of the pQCD
inspired KKMR procedure (for more details see \cite{PST07}).

The situation with the axial-vector production is new compared to
both zero-spin case (scalar \cite{PST07}, pseudoscalar \cite{SPT07}
mesons) as well as to the vector meson production where the vector
meson is dominantly transversely polarized \cite{SS07,RSS08}, at
least, for small transferred four-momenta in the nucleon lines. The
axial-vector meson, as it will be discussed here, can be polarized
both transversely and longitudinally. We shall calculate the cross
section for different polarisation states of the $\chi_c(1^+)$
meson.

There is interesting theoretical aspect of the double diffractive
production of the $\chi_c(1^{+})$ meson. The coupling
$g^*g^*\chi_c(1^{+}) $ (see Fig.~\ref{fig:QCDdiff}) vanishes for
on-shell gluons (so-called Landau-Yang theorem). According to the
original Landau-Yang theorem \cite{LY_theorem} the symmetries under
space rotation and inversion forbid the decay of the spin-1 particle
into two (on-shell) spin-1 particles (two photons, two gluons). The
same is true for the fusion of two on-shell gluons. The symmetry
arguments cannot be strictly applied for off-shell gluons. This fact
has been already explored in inclusive production of $\chi_c(1^+)$
\cite{HKSST01,Saleev06,LL,Baranov}, in the production of spin-1
glueballs \cite{Murgia}, and recently in the studies of decays of
hypothetical $Z'$ bosons into pair of standard $Z$ bosons
\cite{Zprime_decay}. One of the goals of our paper is to confirm
explicitly that the Landau-Yang theorem is violated by virtual
effects in diffractive production of $\chi_c(1^+)$ leading to very
important observational consequences. In our approach the off-shell
effects are treated explicitly. For comparison, in the standard KKMR
approach the corresponding cross section would vanish due to their
on-shell approximation. The measurement of the cross section can be
therefore a good test of the off-shell effects and, consequently,
UGDFs used in the calculation.

At the Tevatron the $\chi_c$ mesons are measured through the $\gamma
+ J/\Psi$ decay channel. The axial-vector $\chi_c(1^+)$ meson has a
large branching fraction of the radiative decay $\chi_c(1^+) \to
\gamma + J/\psi$ (BR = 0.36 \cite{PDG}). This is much bigger than
for the scalar $\chi_c(0^+)$ where it is only about 1 \% \cite{PDG}.
Therefore, the discussed off-shell effects may be very important to
understand the situation in the $\gamma + J/\Psi$ channel observed
experimentally.

\section{Formalism}

The kinematics of the process was already discussed in our previous
paper on $\chi_c(0^+)$ \cite{PST07}. Here we discuss only details of
the matrix element for exclusive $\chi_c(1^+)$ production. This is
derived for the first time, at least in the generalized KKMR
approach.

\subsection{General $\chi_{cJ}$ production amplitude}

In the following we employ the general Kaidalov-Khoze-Martin-Ryskin
approach \cite{KMR,KKMR,KKMR-spin}, and write the amplitude of the
exclusive double diffractive color singlet production $pp\to
pp\chi_{cJ}$ as
\begin{eqnarray}
{\cal
M}^{p p \to p p \chi_{cJ}}_{J,\lambda}=\frac{s}{2}\cdot
\pi^2\frac12\frac{\delta_{c_1c_2}}{N_c^2-1}\,\Im\int
d^2
q_{0,t}V^{c_1c_2}_{J,\lambda}(q_1,q_2,p_M) \;
\nonumber \\
\times\frac{f^{off}_{g,1}(x_1,x_1',q_{0,t}^2,q_{1,t}^2,t_1)
      f^{off}_{g,2}(x_2,x_2',q_{0,t}^2,q_{2,t}^2,t_2)}
{q_{0,t}^2\,q_{1,t}^2\, q_{2,t}^2} \; .
\label{ampl}
\end{eqnarray}
In this expression $f^{off}_{g,i}(x_i,x_i',q_{0,t}^2,q_{i,t}^2,t_i)$
are off-diagonal unintegrated gluon distributions. In the general
case we do not know the off-diagonal UGDFs very well. In
Ref.~\cite{SPT07,PST07} we have proposed a prescription how to
calculate the off-diagonal UGDFs with the help of their diagonal
counterparts:
\begin{eqnarray}\nonumber
f_{g,1}^{off} &=& \sqrt{f_{g}^{(1)}(x_1',q_{0,t}^2,\mu_0^2) \cdot
f_{g}^{(1)}(x_1,q_{1,t}^2,\mu^2)} \cdot F_1(t_1)\,, \\
f_{g,2}^{off} &=& \sqrt{f_{g}^{(2)}(x_2',q_{0,t}^2,\mu_0^2) \cdot
f_{g}^{(2)}(x_2,q_{2,t}^2,\mu^2)} \cdot F_1(t_2)\,,
\label{skewed_UGDFs}
\end{eqnarray}
where $F_1(t_1)$ and $F_1(t_2)$ are the isoscalar nucleon form
factors. In the present work we shall use a few sets of unintegrated
gluon distribution functions (UGDFs), which aim at description of
phenomena where small gluon transverse momenta are involved. Some
details concerning these distributions can be found in
Ref.~\cite{LS06}. We shall follow notations there.

Following our previous work \cite{PST07}, the vertex factor
$V_{J}^{c_1c_2}\equiv V_{J}^{c_1c_2}(q_{1,t}^2,
q_{2,t}^2,P_{M,t}^2)$ in Eq.~(\ref{ampl}) describing the coupling of
two virtual gluons to $\chi_{cJ}$-meson follows from
\begin{eqnarray}
V^{c_1c_2}_J(q_1,q_2)
={\cal P}(q\bar{q}\rightarrow\chi_{cJ})
\bullet\Psi^{c_1c_2}_{ik}(q_1,q_2),
\label{vert}
\end{eqnarray}
where ${\cal P}(q\bar{q}\rightarrow\chi_{cJ})$ is
the operator that projects the $q\bar{q}$ pair onto
the charmonium bound state
(see below), $\Psi^{c_1c_2}(q_1,q_2)$ is the production
amplitude of a pair of massive quark $q$ and antiquark
$\bar{q}$ with momenta $k_1$, $k_2$, respectively.

Within the quasi-multi-Regge-kinematics (QMRK) approach
\cite{FL96} we have
\begin{eqnarray}\label{qqamp}
\Psi(c_1,c_2;i,k;q_1,q_2)&=&
-g^2(t^{c_1}_{ij}t^{c_2}_{jk}b(k_1,k_2)-t^{c_2}_{kj}t^{c_1}_{ji}\bar{b}(k_2,k_1)),\quad
\alpha_s=\frac{g^2}{4\pi},
\end{eqnarray}
where $t^c$ are the colour group generators in the fundamental
representation, $b,\,\bar{b}$ are the effective vertices arising
from the Feynman rules in QMRK
\begin{eqnarray} \label{bb}
b(k_1,k_2)=\gamma^-\frac{\hat{q}_{1}-\hat{k}_{1}-m}{(q_1-k_1)^2-m^2}
\gamma^+,\quad\bar{b}(k_1,k_2)=\gamma^+\frac{\hat{q}_{1}-\hat{k}_{1}+m}{(q_1-k_1)^2-m^2}
\gamma^- \; .
\end{eqnarray}
While projecting on the color singlet the $ggg$-vertex contributions
disappear from the resulting matrix element, so we did not write
them explicitly in Eq.~(\ref{bb}). Taking into account standard
definitions of the light-cone vectors $n^+=p_2/E_{cms},\;
n^-=p_1/E_{cms}$ and momentum decompositions $q_1=x_1p_1+q_{1,t},\;
q_2=x_2p_2+q_{2,t}$ and using the gauge invariance property
(Gribov's trick) one gets the following projection
\begin{eqnarray}
&&\quad q_1^{\nu}V^{c_1c_2}_{J,\,\mu\nu}=
q_2^{\mu}V^{c_1c_2}_{J,\,\mu\nu}=0, \nonumber\\
V^{c_1c_2}_{J}(q_1,q_2)&=&
n^+_{\mu}n^-_{\nu}V_{J,\,\mu\nu}^{c_1c_2}(q_1,q_2)=
\frac{4}{s}\frac{q^{\nu}_{1,t}}{x_1}\frac{q^{\mu}_{2,t}}{x_2}
V^{c_1c_2}_{J,\,\mu\nu}(q_1,q_2). \label{decomp}
\end{eqnarray}
Since we adopt here the definition of the polarization vectors
proportional to gluon transverse momenta $q_{1/2,t}$, then we must
take into account the longitudinal momenta in the numerators of
vertices (\ref{bb}).

Projection of the hard amplitude onto the singlet charmonium bound
state $V_{\mu\nu}^{c_{1}c_{2}}$ is given by the 4-dimensional
integral over relative momentum of quark and antiquark
$q=(k_1-k_2)/2$ \cite{HKSST00,HKSST01}:
\begin{eqnarray}\nonumber
&&V_{J,\,\mu\nu}^{c_{1}c_{2}}(q_1,q_2)={\cal
P}(q\bar{q}\rightarrow\chi_{cJ})\bullet
\Psi^{c_1c_2}_{ik,\,\mu\nu}(q_1,q_2)=
2\pi\cdot\sum_{i,k}\sum_{L_{z},S_{z}}\frac{1}{\sqrt{m}}\int
\frac{d^{\,4}q}{(2\pi )^{4}}\delta \left(
q^{0}-\frac{{\bf q}^{2}}{M}\right)\times\\
&&\times\,\Phi_{L=1,L_{z}}({\bf q})\cdot\left\langle
L=1,L_{z};S=1,S_{z}|J,J_{z}\right\rangle \left\langle
3i,\bar{3}k|1\right\rangle {\rm
Tr}\left\{\Psi_{ik,\,\mu\nu}^{c_{1}c_{2}}{\cal
P}_{S=1,S_{z}}\right\}, \label{amplitude-diff} \\
&&\Psi_{ik,\,\mu\nu}^{c_{1}c_{2}}=-g^2 \;
\sum_j
\biggl[t^{c_1}_{ij}t^{c_2}_{jk}\cdot
\biggl\{\gamma_{\nu}\frac{\hat{q}_{1,t}-\hat{k}_{1,t}-m}
{(q_1-k_1)^2-m^2}\gamma_{\mu}\biggr\}-t^{c_2}_{kj}t^{c_1}_{ji}\cdot
\biggl\{\gamma_{\mu}\frac{\hat{q}_{1,t}-\hat{k}_{2,t}+m}{(q_1-k_2)^2-m^2}
\gamma_{\nu}\biggr\}\biggr].\nonumber
\end{eqnarray}
where the function $\Phi_{L=1,L_{z}}({\bf q})$ is the momentum space
wave function of charmonium, and for a small relative momentum $q$
the projection operator ${\cal P}_{S=1,S_{z}}$ has the form
\begin{eqnarray}
{\cal
P}_{S=1,S_{z}}=\frac{1}{2m}(\hat{k}_2-m)\frac{\hat{\epsilon}(S_{z})}
{\sqrt{2}}(\hat{k}_1+m) \; .
\end{eqnarray}

Since $P$-wave function $\Phi_{L=1,L_{z}}$ vanishes at the origin,
we may expand the trace in Eq.~(\ref{amplitude-diff}) in Taylor
series around ${\bf q}=0$, and only the linear terms in $q^{\sigma}$
in the trace survive. This yields an expression proportional to
\begin{eqnarray}\label{expansion}
\int \frac{d^{3}{\bf
q}}{(2\pi)^{3}}q^{\sigma}\Phi_{L=1,L_{z}}({\bf q})=-i
\sqrt{\frac{3}{4\pi}}\epsilon^{\sigma}(L_{z}){\cal R}^{\prime}(0),
\end{eqnarray}
with the derivative of the $P$-wave radial wave function at the
origin ${\cal R}^{\prime}(0)$ whose numerical values can be found in
Ref.~\cite{EQ95}. The general $P$-wave result (\ref{amplitude-diff})
may be further reduced by employing the Clebsch-Gordan identity
which for the vector $\chi_{cJ=1}$ charmonium reads
\begin{eqnarray} \label{T}
&&{\cal
T}^{\sigma\rho}_{J=1}\equiv\sum_{L_{z},S_{z}}\!\!\left\langle
1,L_{z};1,S_{z}|1,J_{z}\right\rangle
\epsilon^{\sigma}(L_{z})\epsilon^{\rho}(S_{z})
\!=\!-i\sqrt{\frac{1}{2}}\varepsilon^{\sigma\rho\alpha\beta}
\frac{P^{\,\alpha}}{M}\epsilon_{\beta}(J_{z}) \; .
\end{eqnarray}
%

\subsection{$g g \to \chi_c(1^+)$-vertex function}

Summarizing all ingredients above in Eqns.~(\ref{decomp}),
(\ref{amplitude-diff}), (\ref{expansion}) and (\ref{T}), we get the
vertex factor in the following covariant form
\begin{eqnarray}
V^{c_1c_2}_{J=1}&=&2g^2\delta^{c_1c_2}\sqrt{\frac{6}{M\pi N_c}}
\frac{{\cal R}'(0)}{M^2(q_1q_2)^2}\,
\varepsilon_{\sigma\rho\alpha\beta}\epsilon^{\beta}(J_{z})\biggl[
q_{1,t}^{\sigma}q_{2,t}^{\rho}(x_1p_1^{\alpha}-x_2p_2^{\alpha})(q_{1,t}^2+q_{2,t}^2)-
\label{Vgen}\\
&-&\dfrac{2}{s}\,p_1^{\sigma}p_2^{\rho}\,
\biggl(q_{1,t}^{\alpha}(2q_{2,t}^2(q_1q_2)-(q_{1,t}q_{2,t})(q_{1,t}^2+q_{2,t}^2))
-q_{2,t}^{\alpha}(2q_{1,t}^2(q_1q_2)-(q_{1,t}q_{2,t})(q_{1,t}^2+q_{2,t}^2))\biggr)\biggr].
\nonumber
\end{eqnarray}

The general vertex function (\ref{Vgen}) possesses the Bose symmetry
under simultaneous permutation of gluon momenta $q_1\leftrightarrow
q_2$ and polarisation vectors $n^{+} \leftrightarrow n^{-}$ defined
in Eq.~(\ref{decomp}), or, equivalently, under simultaneous
permutations of protons ($p_1\leftrightarrow p_2$) and  gluons (both
transverse $q_{1,t}\leftrightarrow q_{2,t}$ and longitudinal
$x_1p_1\leftrightarrow x_2p_2$) momenta.

We write the decomposition of the polarisation vector of a heavy
meson with a given helicity $\lambda=0,\pm 1$ as
\begin{eqnarray*}
\epsilon^{\,\beta}(P,\lambda)=(1-|\lambda|)n_3^{\beta}-\frac{1}{\sqrt{2}}\,(\lambda
n_1^{\beta}+i|\lambda|n_2^{\beta}),\;\;
n^{\mu}_0=\frac{P_{\mu}}{M},\;\;
n^{\mu}_{\alpha}n^{\nu}_{\beta}g_{\mu\nu}=g_{\alpha\beta},\;\;\epsilon^{\mu}
(\lambda)\epsilon^*_{\mu}(\lambda')=-\delta^{\lambda\lambda'}
\end{eqnarray*}
In the c.m.s. frame we choose the basis with collinear ${\bf n}_3$
and ${\bf P}$ vectors (so, we have ${\bf P}=(E,0,0,P_z),\;P_z=|{\bf
P}|>0$) as a simplest one
\begin{eqnarray}
n_1^{\beta}=(0,\,1,\,0,\,0),\;\; n_2^{\beta}=(0,\,0,\,1,\,0),\;\;
n_3^{\beta}=\frac{1}{M}\,(|{\bf P}|,0,0,E),\;\; |{\bf
P}|=\sqrt{E^2-M^2}. \label{basis}
\end{eqnarray}
%
\begin{figure}[h!]
 \centerline{\epsfig{file=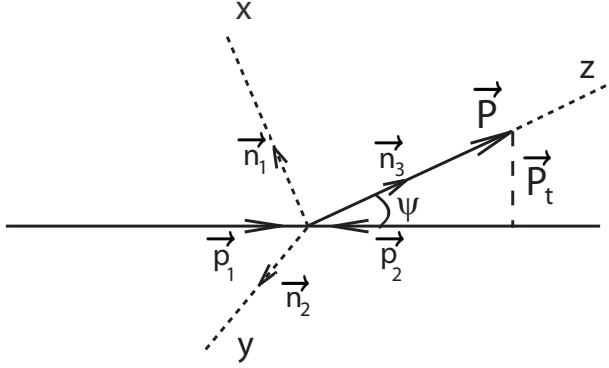,height=5cm,width=8cm}}
 \caption{\footnotesize Coordinate basis in the center-of-mass system
 of incoming protons $p_{1,2}$.}
 \label{fig:cms}  \end{figure}
Note, that we choose ${\bf n}_2$ to be transverse to the c.m.s beam
axis (see Fig.~\ref{fig:cms}), while ${\bf n}_1,\,{\bf n}_3$ are
turned around by the polar angle $\psi=[0\,...\,\pi]$ between ${\bf
P}$ and the c.m.s. beam axis. In the considered basis $\{{\bf
n}_1,\,{\bf n}_2,\,{\bf n}_3\}$ we have the following coordinates of
the incoming protons
\begin{eqnarray}\label{protons}
p_1=\frac{\sqrt{s}}{2}(1,\,-\sin\psi,\,0,\,\cos\psi),\quad
p_2=\frac{\sqrt{s}}{2}(1,\,\sin\psi,\,0,\,-\cos\psi) \; .
\end{eqnarray}

The gluon transverse momenta with respect to the c.m.s. beam axis
are
\begin{eqnarray*}\nonumber
q_{1,t}=(0,\,Q_{1,t}^{x}\cos\psi,\,Q_{t}^y,\,Q_{1,t}^{x}\sin\psi),\quad
q_{2,t}=(0,\,Q_{2,t}^{x}\cos\psi,\,-Q_{t}^y,\,Q_{2,t}^{x}\sin\psi),
\end{eqnarray*}
where $Q_{1/2,t}^{x},\,\pm Q_{t}^{y}$ are the components of the
gluon transverse momenta in the basis with the $z$-axis collinear to
the c.m.s. beam axis.

From definition (\ref{protons}) it follows that energy of the meson
and polar angle $\psi$ are related to covariant scalar products in
the considered coordinate system as
\begin{eqnarray}\label{Epsi}
E=\frac{(p_1P)+(p_2P)}{\sqrt{s}},\quad
\cos\psi=\frac{(p_1P)-(p_2P)}{\sqrt{s}|{\bf P}|},\quad
\sin\psi=\frac{(p_2n_1)-(p_1n_1)}{\sqrt{s}}.
\end{eqnarray}
Further, we also see that from $q_1=x_1p_1+q_{1,t},\;
q_2=x_2p_2+q_{2,t}$ and $q_1+q_2=P$ we have
\begin{eqnarray}\label{x1x2cos}
x_1=\frac{E+|{\bf P}|\cos\psi}{\sqrt{s}},\qquad x_2=\frac{E-|{\bf
P}|\cos\psi}{\sqrt{s}} \; .
\end{eqnarray}
Relations (\ref{Epsi}) and (\ref{x1x2cos}) show that the interchange
of proton momenta $p_1\leftrightarrow p_2$ is equivalent to the
interchange of the angle $\psi\leftrightarrow\psi\pm\pi$, i.e.
$\sin\psi\leftrightarrow-\sin\psi$ and
$\cos\psi\leftrightarrow-\cos\psi$, simultaneously. The last
permutation also provides the interchange of the longitudinal
components of gluons momenta $x_1\leftrightarrow x_2$.

Conservation laws provide us with the following relations between
components of gluon transverse momenta and covariant scalar products
\begin{eqnarray}\nonumber
&&Q_{1,t}^{x}=-\frac{q_{1,t}^2+(q_{1,t}q_{2,t})}{|{\bf
P}|\sin\psi},\quad
Q_{2,t}^{x}=-\frac{q_{2,t}^2+(q_{1,t}q_{2,t})}{|{\bf
P}|\sin\psi},\quad
Q_t^{y}=\frac{\sqrt{q_{1,t}^2q_{2,t}^2-(q_{1,t}q_{2,t})^2}}{|{\bf
P}_t|}\,
\mathrm{sign}(Q_t^y),\label{rel_comp} \\
&&P_{t}^2=-|{\bf P}_t|^2=-|{\bf
P}|^2\sin^2\psi=q_{1,t}^2+q_{2,t}^2+2(q_{1,t}q_{2,t}),\quad
q_{1/2,t}^2=-|{\bf q}_{1/2,t}|^2, \nonumber
\end{eqnarray}
where $|{\bf P}_t|=|{\bf P}||\sin\psi|$ is the meson transverse
momentum with respect to $z$-axis. The appearance of the factor
$\mathrm{sign}(Q_t^y)$ guarantees the applicability of
Eq.~(\ref{rel_comp}) for positive and negative $Q_t^y$. Note that
under permutations $q_{1,t}\leftrightarrow q_{2,t}$ implied by the
Bose statistics the components interchange as
$Q_{1,t}^{x}\leftrightarrow Q_{2,t}^{x}$ and
$Q_{t}^{y}\leftrightarrow -Q_{t}^{y}$. In our notations the quantity
$\sin\psi$ plays a role of the noncollinearity of meson in
considered coordinates. A straightforward calculation leads to the
following vertex function in these coordinates
\begin{eqnarray}\nonumber
V^{c_1c_2}_{J=1,\,\lambda}&=&-8g^2\delta^{c_1c_2}\sqrt{\frac{6}{M\pi
N_c}} \frac{{\cal R}'(0)}{|{\bf P}_t|(M^2-q_{1,t}^2-q_{2,t}^2)^2}
\biggl\{\frac{1}{\sqrt{2}}\biggl[i|\lambda|(q_{1,t}^2-q_{2,t}^2)(q_{1,t}q_{2,t})
\mathrm{sign}(\sin\psi)\\
&+&\lambda(q_{1,t}^2+q_{2,t}^2)\,|[{\bf q}_{1,t}\times{\bf
q}_{2,t}]\times{\bf
n}_1|\,\mathrm{sign}(Q_t^y)\,\mathrm{sign}(\cos\psi)\biggr]+\label{V-fin}\\
&+&(1-|\lambda|)(q_{1,t}^2+q_{2,t}^2)\,|[{\bf q}_{1,t}\times{\bf
q}_{2,t}]\times{\bf
n}_3|\,\mathrm{sign}(Q_t^y)\,\mathrm{sign}(\sin\psi)\biggr\}
\nonumber
\end{eqnarray}
where
\begin{eqnarray*}
&&|[{\bf q}_{1,t}\times{\bf q}_{2,t}]\times{\bf
n}_1|=\sqrt{q_{1,t}^2q_{2,t}^2-(q_{1,t}q_{2,t})^2}\,|\cos\psi|,\\
&&|[{\bf q}_{1,t}\times{\bf q}_{2,t}]\times{\bf
n}_3|=\frac{E}{M}\sqrt{q_{1,t}^2q_{2,t}^2-(q_{1,t}q_{2,t})^2}\,|\sin\psi|.
\end{eqnarray*}
The amplitude (\ref{V-fin}) explicitly obeys the Bose symmetry under
the interchange of gluon momenta and polarisations due to resulting
simultaneous permutations $\cos\psi\leftrightarrow-\cos\psi$,
$\sin\psi\leftrightarrow-\sin\psi$ and $Q_{t}^{y}\leftrightarrow
-Q_{t}^{y}$\footnote{We are thankful to M.~G.~Ryskin for
enlightening correspondence on the issues of Bose symmetry of
production amplitude.}.

A short inspection of Eq.~(\ref{V-fin}) shows that
\begin{equation}
V_{J=1,\lambda}^{c_1,c_2}(q_{1,t},q_{2,t}) \to 0
\end{equation}
when $q_{1,t} \to $ 0 or $q_{2,t} \to $ 0. It shows that gluon
transverse momenta (gluon virtualities) are necessary to get a
nonzero cross section. It also means that the amplitude and the
cross section are sensitive to larger values of gluon transverse
momenta than e.g. in the case of $\chi_c(0^+)$ production.

It follows from the conservation laws that
\begin{eqnarray}\nonumber
q_{1t}+p'_{1t}=-q_{0t},\qquad q_{2t}+p'_{2t}=q_{0t},\qquad
P_t=-(p'_{1t}+p'_{2t})
\end{eqnarray}
Let us consider firstly the limit of the ``coherent'' scattering
protons $p'_{1t}=p'_{2t}\equiv p_t$, so
\begin{eqnarray}\label{coher-mom}
q_{1t}=-(p_t+q_{0t}),\qquad q_{2t}=-(p_t-q_{0t}),\qquad
P_t=-2p_{t},\qquad p_t^y=0.
\end{eqnarray}
The production vertex (\ref{V-fin}) in this limit and considered
coordinates has a form
\begin{eqnarray}\nonumber
V^{c_1c_2}_{J=1,\,\lambda}(q_{0t}^x,q_{0t}^y,p_t)&=&-16g^2\delta^{c_1c_2}\sqrt{\frac{3}{M\pi
N_c}} \frac{{\cal R}'(0)}{(M^2-2(p_t^2+q_{0t}^2))^2}
\biggl\{i|\lambda|(q_{0t}^2-p_t^2)q_{0t}^x\,\mathrm{sign}(\sin\psi)+\\
&+&(p_t^2+q_{0t}^2)q_{0t}^y\bigg[\lambda\cos\psi+\frac{\sqrt{2}E}{M}(1-|\lambda|)\sin\psi\bigg]\biggr\}.
\label{coher}
\end{eqnarray}
This vertex is antisymmetric w.r.t. simultaneous interchanges
$q_{0t}^x\leftrightarrow-q_{0t}^x$ and
$q_{0t}^y\leftrightarrow-q_{0t}^y$:
\begin{eqnarray}
V^{c_1c_2}_{J=1,\,\lambda}(q_{0t}^x,q_{0t}^y,p_t)=-V^{c_1c_2}_{J=1,\,\lambda}(-q_{0t}^x,-q_{0t}^y,p_t)
\end{eqnarray}
In the considered
``coherent'' limit (\ref{coher-mom}) the integrand of the diffractive
amplitude 
\begin{eqnarray*}
\frac{V^{c_1c_2}_{J=1,\,\lambda}(q_{0t}^x,q_{0t}^y,p_t)\cdot
f^{off}_{g,1}(x_1,x_1',q_{0,t}^2,(p_t+q_{0t})^2,t_1)
f^{off}_{g,2}(x_2,x_2',q_{0,t}^2,(p_t-q_{0t})^2,t_2)}{q_{0t}^2(p_t+q_{0t})^2(p_t-q_{0t})^2}
\end{eqnarray*}
will be antisymmetric only if $x_1=x_2=E/\sqrt{s}\equiv x$ (while
$x_1'\sim x_2'\ll x_{1,2}$), i.e. in the case when $y=0$, while the deviation from zero at 
$y \not=0$ manifests the violation of Regge factorization which was used to examine this limit  
earlier \cite{KKMR-spin}. So, the
diffractive amplitude in this case
\begin{eqnarray*}
{\cal M}_{y\to0}\sim F_1(t_1)F_1(t_2)\int
dq_{0t}^xdq_{0t}^y\frac{V_{J=1}(q_{0t}^x,q_{0t}^y,p_t)\cdot
f(x,q_{0,t}^2,q_{1,t}^2)f(x,q_{0,t}^2,q_{2,t}^2)}{q_{0t}^2q_{1t}^2q_{2t}^2}=0.
\end{eqnarray*}

In the forward limit $p_{t}\to0$ (which is the particular case of coherent one) 
the amplitude turns to zero at any $y$. Indeed, 
we have $P_t\to0$ and
$\sin\psi\to\pm0$ and the amplitude turns into
\begin{eqnarray} \label{pt0}
V^{c_1c_2}_{J=1,\,\lambda}(q_{0t}^x,q_{0t}^y,p_t\to0)&=&-16g^2\delta^{c_1c_2}{\cal
R}'(0)\sqrt{\frac{3}{M\pi
N_c}}\times\\
&\times&\frac{q_{0t}^2}{(M^2-2q_{0t}^2)^2}\bigg\{i|\lambda|q_{0t}^x\,
\mathrm{sign}(\sin\psi)|_{\psi\to0,\pi}+\lambda
q_{0t}^y\,\mathrm{sign}(\cos\psi)|_{\psi\to0,\pi}\bigg\} \nonumber
\end{eqnarray}
As in the previous case, it is obviously antisymmetric under
interchanges $q_{0t}^x\leftrightarrow-q_{0t}^x$ and
$q_{0t}^y\leftrightarrow-q_{0t}^y$. Since in this case
$q_{1t}=-q_{0t},\,q_{2t}=q_{0t}$, then the diffractive amplitude
has an antisymmetric integrand and turns to zero
\begin{eqnarray*}
{\cal M}_{p_t\to0}\sim F_1(t_1)F_1(t_2)\int
dq_{0t}^xdq_{0t}^y\frac{V_{J=1}(q_{0t}^x,q_{0t}^y,p_t\to0)\cdot
f(x_1,q_{0,t}^2,q_{0,t}^2)f(x_2,q_{0,t}^2,q_{0,t}^2)}{q_{0t}^6}=0.
\end{eqnarray*}
This explicitly confirms the observation made in
Refs.~\cite{KKMR-spin,Yuan01}\footnote{We are grateful to
V.~A.~Khoze for very interesting and helpful correspondence on this problem.}.

It is also possible to
express the results in terms of transverse 3-momenta of fusing off-shell
gluons $|{\bf q}_{1,t}|$ and $|{\bf q}_{2,t}|$, and the angle
between them $\phi$ in the center-of-mass system of colliding
nucleons with $z$-axis fixed along meson momentum ${\bf P}$. In this
case, summing the squared matrix elements over meson polarizations we
get the expression 
\begin{eqnarray}\nonumber
\frac{|{\bf
q}_{1,t}|^2|{\bf q}_{2,t}|^2 \Big[\left(|{\bf q}_{1,t}|^2+|{\bf
q}_{1,t}|^2\right)^2\sin^2\phi+M^2\left(|{\bf q}_{1,t}|^2+ |{\bf
q}_{2,t}|^2-2|{\bf q}_{1,t}||{\bf q}_{2,t}|\cos\phi\right)\Big]}
{(|{\bf q}_{1,t}|^2+|{\bf q}_{2,t}|^2+M^2)^4},
\label{tot-sum}
\end{eqnarray}
equal (up to different normalisations of gluon
polarization vectors)  to the one derived in Ref.~\cite{Saleev06}.

\subsection{Three-body phase space}

At high energies and small momentum transfers the phase space
volume element can be written as \cite{KMV99}
\begin{eqnarray}
d^3 PS = \frac{1}{2^8 \pi^4} dt_1 dt_2 d\xi_1 d\xi_2 d \Phi \;
\delta \left( s(1-\xi_1)(1-\xi_2)-M^2 \right) \; ,
\label{dPS_element_he1}
\end{eqnarray}
where $\xi_1$, $\xi_2$ are longitudinal momentum fractions carried
by outgoing protons with respect to their parent protons and the
relative angle between outgoing protons $\Phi \in (0, 2\pi)$.
Changing variables  $(\xi_1, \xi_2) \to (x_F, M^2)$ one gets
\begin{eqnarray}
d^3 PS = \frac{1}{2^8 \pi^4} dt_1 dt_2 \frac{dx_F}{s \sqrt{x_F^2 + 4
(M^2+|{\bf P}_{M,t}|^2)/s}} \; d \Phi \; .
\label{dPS_element_he2}
\end{eqnarray}

\section{Results}

\begin{figure}[!h]    
\includegraphics[width=0.4\textwidth]{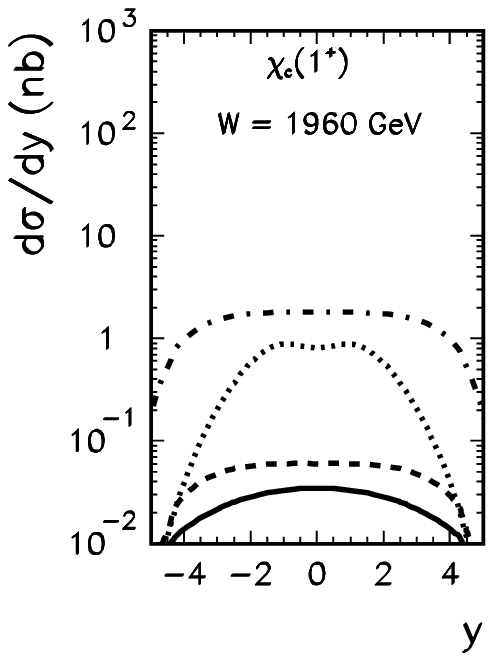}
\hspace{1cm}
\includegraphics[width=0.4\textwidth]{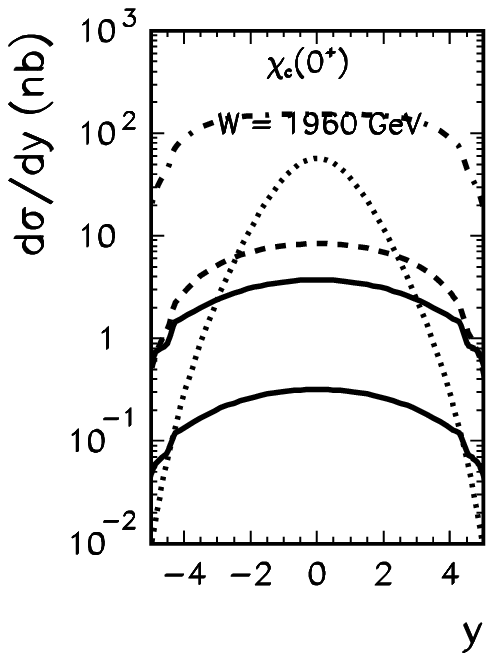}
   \caption{\label{fig:dsig_dy}
   \small Distributions in rapidity of $\chi_c(1^+)$ meson
(left panel) and $\chi_c(0^+)$ meson (right panel) for different
UGDFs. Dash-dotted line corresponds to BFKL UGDF, long-dashed line
-- GBW, short-dashed line -- KL, and two solid lines -- Gaussian
UGDFs for $\sigma_0=0.5$ GeV$^2$ (upper line) and $\sigma_0=1.0$
GeV$^2$ (lower line).}
\end{figure}

Let us start from presenting the differential cross sections. In
Fig.~\ref{fig:dsig_dy} we show distributions in rapidity $y$ for
different UGDFs from the literature. The results for different UGDFs
significantly vary. The biggest cross section is obtained with BFKL
UGDF and the smallest one with Gaussian UGDFs. The big spread of the
results is due to quite different distributions of UGDFs in gluon
transverse momenta $q_{1t}, q_{2t}$, although when integrated over
transverse momenta distributions in longitudinal momentum fractions
$x_1, x_2$ are fairly similar.

Comparing the left and right panels, the cross section for the
axial-vector $\chi_c(1^+)$ production is much smaller (more than an
order of magnitude) than the cross section for the scalar
$\chi_c(0^+)$ production. This is related to the Landau-Yang
theorem, which ``causes'' vanishing of the cross section for
on-shell gluons. For axial-vector quarkonia the effect is purely of
off-shell nature and is due to the interplay of the off-shell matrix
element and off-diagonal UGDFs. This interplay causes a huge
sensitivity of differential distributions to UGDFs observed in
Fig.~\ref{fig:dsig_dy}.
\begin{figure}[!h]    
\includegraphics[width=0.4\textwidth]{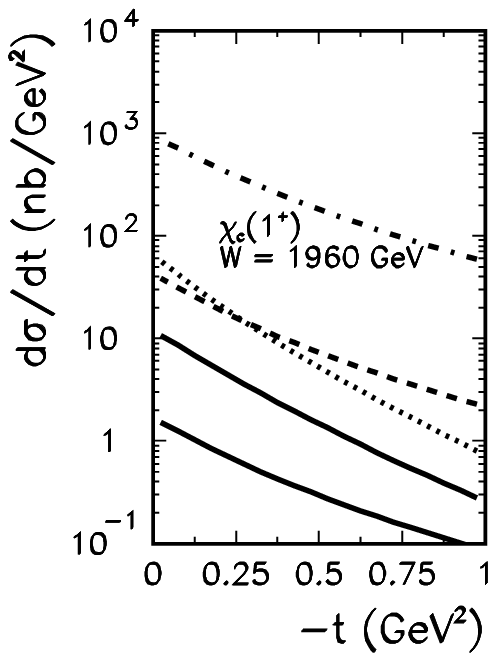}
\hspace{1cm}
\includegraphics[width=0.4\textwidth]{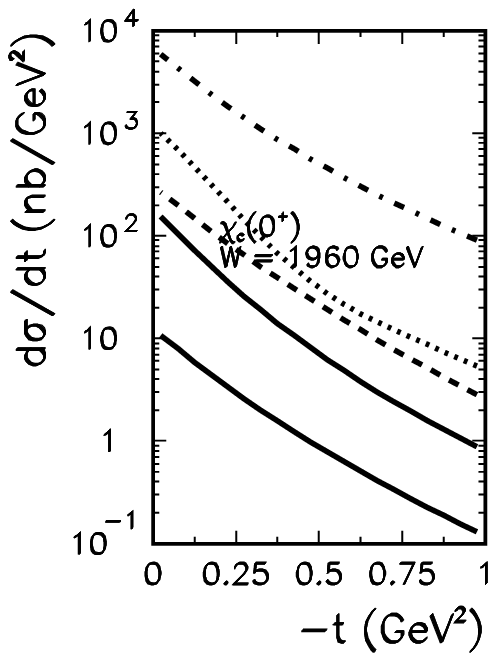}
   \caption{\label{fig:dsig_dt}
   \small Distribution in $t_{1,2}$ of $\chi_c(1^+)$ meson
(left panel) and $\chi_c(0)$ meson (right panel) for different
UGDFs.}
\end{figure}

In Fig.~\ref{fig:dsig_dt} we show corresponding distributions in $t
= t_1$ or $t = t_2$ (identical) again for different UGDFs. Except of
normalisation the shapes are rather similar. This is because of the
$t_1$ and $t_2$ dependencies of form factors, describing the
off-diagonal effect, taken the same for different UGDFs.

\begin{figure}[!h]    
\includegraphics[width=0.4\textwidth]{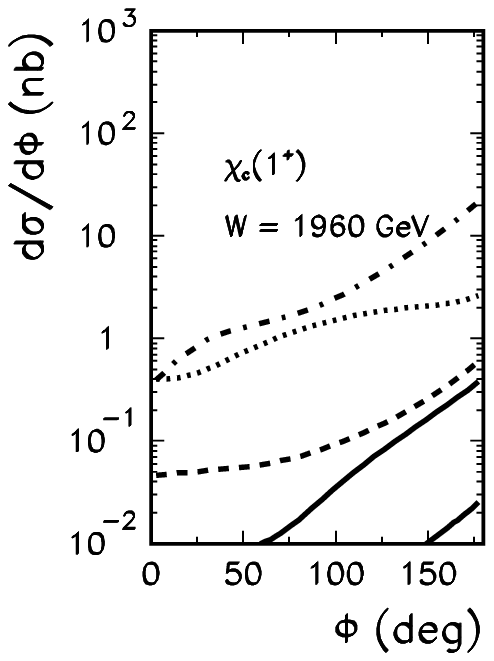}
\hspace{1cm}
\includegraphics[width=0.4\textwidth]{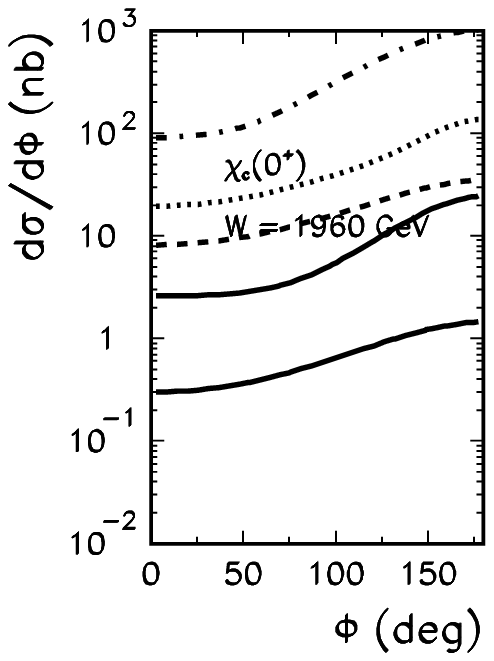}
   \caption{\label{fig:dsig_dphi}
   \small Distribution in relative azimuthal angle $\Phi$ of
$\chi_c(1^+)$ (left panel) and $\chi_c(0^+)$ (right panel) meson
production for different UGDFs.}
\end{figure}

In Fig.~\ref{fig:dsig_dphi} we show the correlation function in
relative azimuthal angle between outgoing protons. The shapes of the
distributions are almost independent of UGDFs.
In the case when energy resolution is not enough to separate
contributions form different states of $\chi_c$ ($\chi_c(0^+)$,
$\chi_c(1^+)$, $\chi_c(2^+)$) the distribution in relative azimuthal
angle may, at least in principle, be helpful.

Summarizing differential distributions, the cross sections
(especially their absolute normalisation) strongly depend on the
model of UGDF. In spite of the huge uncertainty in predicting the
absolute cross section it becomes obvious that the cross section for
$\chi_c(0^+)$ is much bigger than the cross section for the
$\chi_c(1^+)$ production. This result could be expected based on the
Landau-Yang theorem. However, the size of the suppression cannot be
predicted without actual calculations.

In Table 1 we have collected cross sections integrated in $y, t_1,
t_2, \phi$ over the full phase space for the Tevatron energy $W$ =
1960 GeV. More than an order of magnitude suppression of
$\chi_c(1^+)$ relative to $\chi_c(0^+)$ can be seen by comparing
numbers in appropriate columns\footnote{This ratio should be only
weakly modified by absorption effects.}.
\begin{table}[!h]
\caption{Integrated cross section $\sigma_{tot}$ (in nb) for
exclusive $\chi_c(0^+)$ and $\chi_c(1^+)$ production for different
UGDFs and the Tevatron energy W = 1960 GeV. Branching ratios of
radiative decays were taken from \cite{PDG_new}: BR$(\chi_c(0^+)\to
J/\Psi\,\gamma)=0.0128$ and BR$(\chi_c(1^+)\to
J/\Psi\,\gamma)=0.36$.}
\begin{center}
\begin{tabular}{|c||c|c||c|c||c|}
\hline
 &\multicolumn{2}{l||}{$\qquad\;\chi_c(0^+)$}&
  \multicolumn{2}{l||}{$\qquad\;\chi_c(1^+)$} & ratio \\
\cline{2-5}
 UGDF&$\quad\sigma_{tot}\quad$&$\;$BR$\cdot\sigma_{tot}\;$&
 $\quad\sigma_{tot}\quad$&$\;$BR$\cdot\sigma_{tot}\;$&
 $\frac{BR\cdot\sigma_{tot}(\chi_c(1^+))}{BR\cdot\sigma_{tot}(\chi_c(0^+))}$ \\
\hline\hline
KL                   & 55.2 & 0.7  & 0.5  & 0.2  & 0.3 \\
GBW                  & 160  &  2   & 4.2  & 1.5  & 0.8 \\
BFKL                 & 1200 & 15.4 & 14.2 & 5.1  & 0.3 \\
Gauss,               &      &      &      &      &     \\
$\sigma_0=0.5$ GeV & 26   & 0.3  & 0.2  & 0.09 & 0.3 \\
Gauss,               &      &      &      &      &     \\
$\sigma_0=1.0$ GeV & 2.2  & 0.03 & 0.02 &0.006 & 0.2 \\
 \hline
\end{tabular}
\end{center}
\end{table}

The best method to measure $\chi_c$ mesons at the Tevatron is via
$\gamma + J/\Psi$ decay channel. All P-wave $\chi_c$-quarkonia decay
into this channel. However, the branching fractions to this channel
are very different \cite{PDG}. While the branching fraction for
$\chi_c(0^+)$ is very small (of the order of 1 \%), the branching
fraction for $\chi_c(1^+)$ is one and half order of magnitude
larger. In the third and fifth columns we present the total cross
sections multiplied by the appropriate branching fractions. After
multiplying the cross section by the branching fraction for $\gamma
+ J/\Psi$ decay the situation somewhat changes, i.e. now
$\chi_c(1^+)$ becomes closer to $\chi_c(0^+)$ --
BR$\cdot\sigma_{tot}(\chi_c(1^+))$ is about two times smaller than
that for $\chi_c(0^+)$ for all UGDFs used in our calculation, so it
seems to be almost model independent statement.

In a preliminary analysis the CDF collaboration
\cite{Albrow_diffraction2008} assumes that the observed strength
comes dominantly from $\chi_c(0^+)$, thus conforming results of our
investigation. In order to make comparison with the experimental
results one would still need to include experimental cuts on lepton
and photon rapidities and transverse momenta. Also including
absorption effects may be important as slightly larger absorption
can be expected for $\chi_c(1^+)$ (harder distributions in $t_1$ and
$t_2$ -- see Fig.~\ref{fig:dsig_dt}). These points need further
studies.

Our calculation suggests that the inclusion of $\chi_c(1^+)$ in the
experimental analysis is not negligible and may be necessary. The
present energy resolution does not allow for separating different
P-wave quarkonia. Perhaps, looking to other decay channels may help
in disentangling the contributions from different states and
allowing for extracting the cross sections separately for each of
them.

Another interesting option to shed more light to the problem is to
study the angular distributions of outgoing $J/\psi$ in the $\chi_c$
rest frame. Different states should have, in principle, different
distributions. We leave the analysis of those distributions for a
separate study.
\begin{table}[!h]
\caption{Integrated cross section $\sigma_{tot}$ (in nb)
for exclusive $\chi_c(1^+)$ production at different
energies.}
\begin{center}
\begin{tabular}{|c|c|c|c|}
\hline
UGDF &$\quad$RHIC$\quad$&$\quad$Tevatron$\quad$& $\quad$LHC$\quad$\\
\hline
KL                   & 0.05  & 0.5   & 1.7   \\
GBW                  & 0.04  & 4.2   & 73.1  \\
BFKL                 & 0.07  & 14.2  & 1064  \\
Gauss,               &       &       &       \\
$\sigma_0=0.5$ GeV   & 0.007 & 0.2   & 2.5   \\
Gauss,               &       &       &       \\
$\sigma_0=1.0$ GeV   & 0.0005 & 0.02 & 0.2   \\
\hline
\end{tabular}
\end{center}
\end{table}

Finally in Table 2 we present the total cross sections
for $\chi_c(1^+)$ also for RHIC and LHC energies.
The question of separation of different $\chi_c$ states
should be similar, except that other decay channels
should be available \cite{Guryn,Schicker}.

\section{Conclusions and discussion}

Our results can be summarized as follows:

We have derived the QCD amplitude for exclusive elastic double
diffractive production of axial-vector $\chi_c(1^+)$ meson.
According to the Landau-Yang theorem the amplitude vanishes for the
fusion of on-shell gluons. In the present analysis we have
generalized the formalism proposed recently for diffractive
production of the Higgs boson. We have derived corresponding $g^*
g^* \to \chi_c(1^+)$ vertex function. Our effect is purely off-shell
type, i.e. requires off-shell gluons, which demands nonvanishing
transverse momenta of gluons in the high-energy regime.

We have calculated the corresponding differential cross sections.
Different unintegrated gluon distributions from the literature have
been used. The absolute cross section is very sensitive to the
choice of UGDF in contrast to the shapes of distributions. The
predicted total (integrated over phase space) cross section,
obtained from the bare amplitude, is from a fraction to several
nanobarns, depending on the model of UGDFs. This is one and a half
order of magnitude less than a similar cross section for
$\chi_c(0^+)$ \cite{PST07}. This is a direct consequence of the
Landau-Yang theorem. However, because the branching fraction
$BR(\chi_c(1^+) \to J/\psi + \gamma) \gg BR(\chi_c(0^+) \to J/\psi +
\gamma)$, one may expect a different situation in the $J/\psi +
\gamma$ channel. This has an analogy with the inclusive production
of P-wave quarkonia, where the signal (in the $J/\psi + \gamma$
channel) of $\chi_c(1^+)$ is larger than that for $\chi_c(0^+)$. We
have observed that BR$\cdot\sigma_{tot}(\chi_c(1^+))$ is (only) several
times smaller than that for $\chi_c(0^+)$ for all UGDFs used in
our calculation.

Moreover, we have also calculated \cite{tobe} differential cross
sections for different spin polarizations of $\chi_c(1^+)$. The
integrated cross section for spin polarization $\lambda=\pm1$ is
approximately an order of magnitude greater than that for the
$\lambda=0$ polarization. Similar observation has already been made
in Refs.~\cite{KKMR-spin,Close} and verified by the WA102 data for
$f_1(1285),\,f_1(1420)$ production \cite{Barberis}. The ratio of the
cross sections integrated over the phase space is only weekly
dependent on UGDFs but strongly depends on $t_1$ and $t_2$.

In the present analysis we have neglected the absorption effects.
The latter clearly go beyond the scope of the present analysis. At
the Tevatron energies they lead, however, to a large damping of the
cross section. In zeroth approximation they can be taken into
account by multiplying the cross section by a so-called soft
survival probability \cite{KMR,KKMR}. At the Tevatron energies the
soft survival probability is of the order of 0.1 \cite{KMR,KKMR}. A
better approximation is to convolute the bare amplitude with
nucleon-nucleon elastic (re)scattering amplitude (see e.g.
Refs.~\cite{SS07,RSS08,PRSG04}). We leave the inclusion and
discussion of the absorption/rescattering effects for a separate
analysis, including all quarkonium states ($\chi_c(0^+),
\chi_c(1^+), \chi_c(2^+)$).

\section{Acknowledgments}

Useful discussions and helpful correspondence with Mike Albrow,
Sergey Baranov, W{\l}odek Guryn, Valery Khoze, Francesco Murgia,
Mikhail Ryskin and Wolfgang Sch\"afer are gratefully acknowledged.
This study was partially supported by the polish grant of MNiSW N
N202 249235, the Russian Foundation for Fundamental Research, grants
No. 07-02-91557 and No. 09-02-01149.



\end{document}